# Evidence of Neutrino Flux effect on Alpha Emission Radioactive Half-Life

Jonathan Walg, Anatoly Rodnianski and Itzhak Orion

Nuclear Engineering, Ben Gurion University of the Negev, Beer Sheva, Israel

## Abstract

Radioactive sources presented annual periodical half-life changes in several accurate measurements, although customary practice claims that radioactive decay should be a physical constant for each radionuclide. Besides that, the Purdue measurements of Mn-54 decay-rates indicated response to solar X-ray flare events in 2006.

In order to track more radiation count-rate responses to solar flare events, we built an experimental detector system for gamma radiation count-rates measurements, facing an Am-241 radioactive source. The system was tracking gamma rays that follow an alpha particle emission. The system was placed at an underground laboratory, permanently locked to avoid any influence by unexpected radiation perturbations, and environmentally controlled in means of temperature and clean-air flow, in order to maintain detectors stabilization. The detector was consists of NaI(Tl) scintillator and total-counting reader devices for remote counting. The radiation counting system was shielded by a 5 cm lead.

One month prior to flare events from the Sun, all three detectors showed reasonably stable count-rates, which were tallied every 15 minutes. Five solar-flares occurred and reported by the SpaceWeatherLive website on 12th to 13th of October 2018. The Am-241 system response to solar flares found to be with a delay of around 16 days, however its response shape is in high significance.

We conclude that also for alpha emitter radioactive sources, the half-life altered due to changes of neutrino flux from the Sun.

Several geological radiometric dating methods, such as the U-235 chain, U-238 chain, and Th-232 chain, that emit alpha particles are in use. Our measurements indicated that an alpha emitter was affected by the neutrino flux change from the Sun. Our new findings should question the reliability of these dating methods since the solar activity varies throughout time as the Sun has been going through an evolutionary process.

**Keywords:** sun, flare, neutrino, radioactivity, half-life.

**Introduction**

Solar x-ray flares occur when the Sun's activity rises, and it is evident that an 11-year sunspot cycle is related to solar activity, therefore there is a higher probability of solar x-ray flares occurring in the higher solar activity phase of the cycle (Solanki, 2003. Now we are at the lowest phase of the solar activity cycle, and although solar flare appearance cannot be accurately predicted, maximal solar activity should appear during the years 2024-2025. The solar X-ray flare phenomenon is thought to be related to the particle transfer loop from the Sun to the corona (Zhu, 2018). In addition, since solar X-ray flares can interact with Earth's ionosphere, several satellites have been launched in order to measure these flares and to report their appearance time and magnitude. A series of GOES (Geostationary Operational Environmental Satellites) satellites operated by the Space Weather Prediction Center, National Oceanic and Atmospheric Administration provides measured solar x-ray flux daily data, which is reported in units of $W/m^2$ for each minute. This x-ray flux daily data is classified as A, B, C, M, or X according to peak flux magnitude, where class A, the lowest flux, is less than $10^{-7}$ $W/m^2$, X is above $10^{-4}$ $W/m^2$, and the difference from class to class is 10-fold.

In previously published research, measurements of half-life radioactive sources presented an annual periodical change, despite the customary notion that radioactive decay should be considered a physical constant for each radionuclide (Lapp, 1954). The most significant publication is the Alburger et al. (1986) experiment (Alburger, 1986), in which decay rates of Si-32 and Cl-36 were simultaneously measured using the same detector system, and annual variations of count-rates were observed to differ in both amplitude and phase. These two radioisotopes are betta emitters, thus the detector's internal response could not be the reason for the counting change. Hence Alburger et al. concluded that half-life varies due to an annual periodical effect.

Since then, further experiments have revealed these half-life annual periodical variations; however, all radioisotopes involved were betta and/or gamma emitters (Jenkins, 2009; Parkhomov, 2011). Yet one recent publication (Sturrock, 2018) presents long-term (i.e., 10 years with 15-minute intervals) measurements of Rn-222 decay data analyzed using spectrograms of the measured gamma radiation followed by the Rn-222 alpha particle emission. Their work proved that Rn-222 alpha particle emission can present an annual periodical count-rate change.

Only one report has been published regarding the influence of solar flares on radioactive half-life (Jenkins, 2009). In December of 2006, for the first time, high-flux x-ray flares (class X – M) were found to be correlated to measured Mn-54 gamma radiation count-rate discrepancies. Mn-54 is an electron-capture radioactive nucleus that produces gamma rays emitter, excited Cr-54, with a 312-day half-life (Firestone, 1999). The hypothesis that solar neutrino flux variations cause these count-rate discrepancies was presented by Jenkins and Fischbach (Jenkins, 2009; Mohsinally, 2016); although assumed to cause these decay rate variations, the involvement of neutrinos in radioactive decay is not included in nuclear physics models. This hypothesis is relaying on neutrino involvement in nuclear decay belongs only to betta decay since it is a lepton that interacts under the weak nuclear field. In addition, it was found that alpha decay rate is dependent on neutrino flux change, therefore neutrinos should also be assumed to interact with the strong nuclear field (Sturrock, 2018).

**Methodology**

Radiation measurement system was integrated in an underground laboratory, one facing an Am-241 (37 kBq) source with a NaI(Tl) detector (2" diameter by 2" length). The detector was shielded with a 5-cm thickness of lead.

Figure 1: Schematic description of the experimental system.

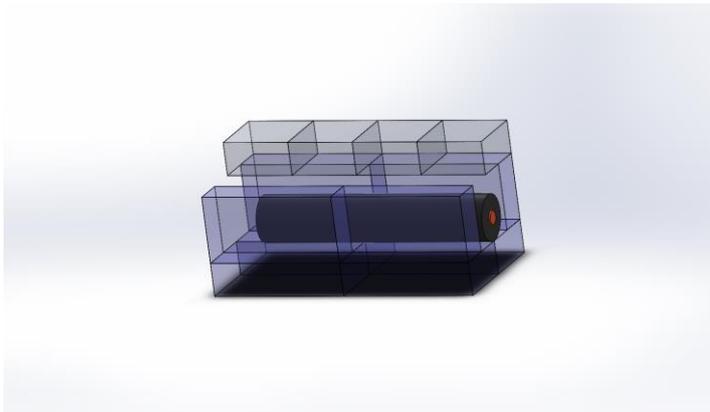

The Am-241 emits a single dominant gamma-ray line of 59.5 keV, as result of alpha decay yields, as illustrated in Fig. 2 (the 43.4 keV line yield is very low compared to the 59.5 keV yield).

Figure 2: The Am-241 decay scheme: alpha emission and gamma rays emission are presented (Firestone, 1999).

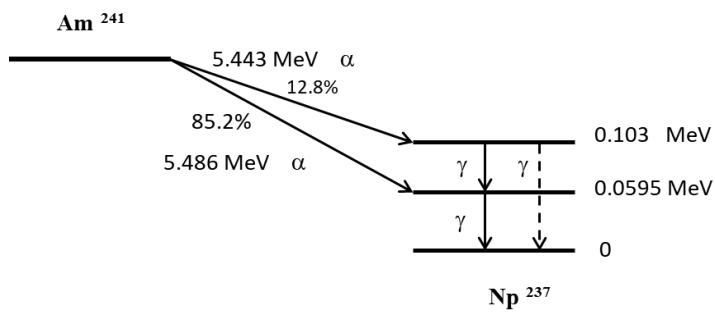

In Fig.1 is the schematic description of the system. The detector was connected to data logger (DL) CR800 manufactured by Campbell Scientific in order to remotely collect and submit data to a computer, which itself is remotely controlled for access to the DL and collected data. Every 15 minutes gamma counts from the detector was integrated and tallied. The lab was permanently locked to avoid any influence and unexpected radiation perturbations, as well as was environmentally controlled in terms of temperature and clean-air flow in order to reduce detector efficiency dependence.

Since August 22th, 2018 the PM-11 has been counting with a 15-minute repetition rate, showing a stable count-rate until October 12th, 2018.

Laboratory temperature, measured throughout the experiment, was found to be stable at 18°C (±1°). Careful temperature stability is required for such delicate changes since scintillation efficiency can be affected by temperature differences (Dolev, 2008), even though peak counts are much more sensitive to these temperature changes compared to total-counts (Knoll, 2000). PM-11 background counts, measured for a two-day period, remained around the level of 700 cpm. Am-241 counting stability was recorded for 150 hours and showed 0.09% counting uncertainty, which we then set as the statistical fluctuations level for the second system, not dependent on temperature variation.

Solar flares were traced on a daily basis from the SpaceWeatherLive website [https://www.spaceweatherlive.com/en/solar-activity/solar-flares], intended for uses related to astronomy, space, space-weather, aurora, etc. This website reports flare details as well as presents their plots sorted by flare intensities and time.

**Results and Discussion**

Five solar flare events of different flux magnitudes occurred between October 12th and October 13th; these flare flux in $nW/m^2$ at the beginning, ending, and maximum times are listed in Table 1.

Table 1: Solar flare occurrence times (UTC) and their flux during October, 2018. Flux accuracy is within 20 $nW/m^2$.

| Date | Start Flare Time UTC | Start Flux ($nW/m^2$) | Max Flare Time UTC | Max Flux ($nW/m^2$) | End Flare Time UTC | End Flux ($nW/m^2$) |
|---|---|---|---|---|---|---|
| 12.OCT.2018 | 01:43 | 25.7 | 01:50 | 220 | 02:17 | 41.8 |
|  | 02:17 | 41.8 | 02:30 | 52.9 | 02:44 | 27.1 |
|  | 14:05 | 29.7 | 14:08 | 611 | 14:45 | 16.0 |
|  | 15:01 | 22.6 | 15:24 | 72.4 | 16:07 | 23.4 |
| 13.OCT.2018 | 13:27 | 12.5 | 13:35 | 239 | 14:02 | 38.7 |

Source: SpaceWeatherLive

Figure 3: Solar flares on October 12th-13th, 2018: full images from SpaceWeatherLive (with permission).

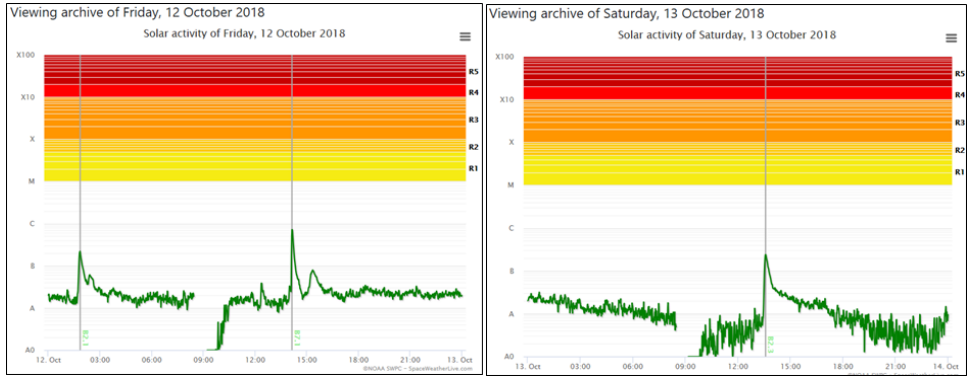

Source: https://www.spaceweatherlive.com/en/solar-activity/solar-flares

Full images from SpaceWeatherLive of the solar flares observed during this study are presented (with permission) in Figure 3. The horizontal axis is UTC time; vertically, solar flux is represented by a log scale classification (A0 – X). The highest flares were measured between classes B and C.

Three count-rate dips are shown in Figure 4 (see red arrows); in the upper-most part of the figure is a graph of these solar flares after adaptation to the UTC+3 time-zone was made.

Figure 4: 59.5 keV gamma radiation of Am-241 count-rates (in 15-min intervals). The blue line is the average cpm obtained prior to the valleys.

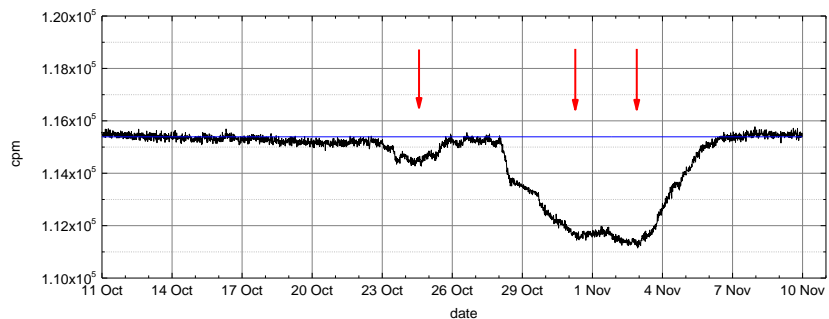

The Am-241 gamma radiation measurements taken during these two days demonstrated a constant count-rate with a mean value of 1.155E+05 cpm (SD of 0.09%); however, a clear valley was evinced on October $28^{th}$ through November $6^{th}$, 2018, at least 16 days following the last solar flare, as shown in Figure 4. These measurements clearly not influenced by external environmental conditions for at least fourteen days, thus these fluctuations are an established fact not only due to the statistical nature of alpha particle decay process, but also to the counting route that might include scintillation, photocathode-electron-emission, and photomultiplier erraticism. A smaller valley taking place on October 24-$25^{th}$, whose minimum is about 0.86% below the average cpm is also shown, leading us to presume that a physical signal was detected.

The first dip in Figure 4 corresponds to the first, relatively short solar flare that occurred on October 12th, and the next valley is the response to the remainder of the solar flares listed in Table 1.

Our flares are orders of magnitude lower compared to the first observed flare in 2006.

Next dips group appeared on December 2018 to January 2019 as shown in Figure 5. These dips corresponded to flares during the beginning of January as listed in Table 2.

Table 2: Solar flare occurrence times (UTC) and their flux from December 2018 to January, 2019

| Date | Start Flare Time UTC | Start Flux (nW/m$^2$) | Max Flare Time UTC | Max Flux (nW/m$^2$) | End Flare Time UTC | End Flux (nW/m$^2$) |
|---|---|---|---|---|---|---|
| 9.DEC.2018 | 07:05 | 27.2 | 07:10 | 243 | 08:15 | 58.3 |
| 15.DEC.2018 | 13:40 | 16.8 | 13:53 | 107 | 14:20 | 24.6 |
| 2.JAN.2019 | 01:40 | 22.0 | 01:50 | 283 | 01:15 | 28.6 |
|  | 18:25 | 30.9 | 18:40 | 121 | 19:00 | 45.7 |
| 3.JAN.2019 | 04:25 | 40.7 | 05:20 | 118 | 07:00 | 33.7 |
|  | 16:25 | 19.4 | 17:15 | 193 | 18:50 | 24.6 |
| 4.JAN.2019 | 10:50 | 16.7 | 11:35 | 428 | 12:20 | 22.8 |
| 6.JAN.2019 | 05:32 | 106 | 05:50 | 860 | 08:20 | 66.5 |
|  | 10:10 | 42.5 | 10:50 | 1660 | 13:15 | 48.9 |
|  | 13:30 | 41.8 | 13:37 | 160 | 13:40 | 40.5 |

Source: SpaceWeatherLive

Figure 5: Gamma radiation of Am-241 count-rates (in 15-min intervals) during December 2018 – January 2019. The blue line is the average cpm obtained prior to the valleys.

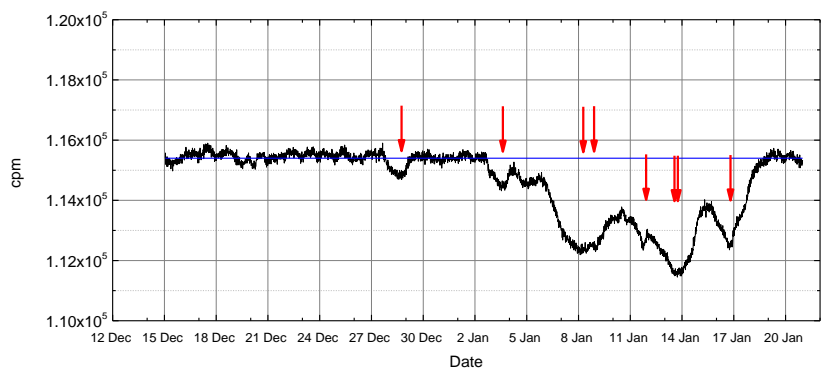

## Conclusions

We found that also for alpha emitter radioactive sources, the half-life altered due to changes of neutrino flux from the Sun. Our measurements, showing more than 10 count dips, indicated that an alpha emitter was affected by the neutrino flux change from the Sun.

The neutrino-flavor-oscillations phenomenon is well-known, as for the neutrino has a finite mass (Barger, 2012). Each neutrino type differs by mass: electron-neutrino is ~ 2 eV, and muon-neutrino mass is ~ $10^5$ higher than the electron-neutrino mass. Along a certain distance the neutrinos are alternating, and due to relativistic momentum-energy equilibrium the electron-neutrino reduces its speed while converting to muon-neutrino (or to tau-neutrino). Therefore, it cannot be assumed that neutrinos travel similarly to speed-of-light from the Sun to the Earth.

Radioactive nucleus that captures a free neutrino into its system may possess it for a certain period until it will affect the count rate alteration. This phenomenon indicates that new nuclear models should be considered for the alpha decay process, as well as that the neutrino be included in internal nucleon structure.

## Acknowledgments

The authors thank the SpaceWeatherLive website team for allowing use of the images in Figure 3.
Our thanks to GSI (Geological Survey Israel) for loaning lab equipment for the sake of this research.